\documentclass[aip,jap,reprint]{revtex4-1}
\usepackage{graphicx}% Include figure files
\usepackage{dcolumn}% Align table columns on decimal point
\usepackage{bm}% bold math

\begin{document}
\title{High pressure effect on structure, electronic structure and thermoelectric properties of MoS$_2$}

\author{Huaihong Guo}
\author{Teng Yang}
\email[E-mail: ]{yangteng@imr.ac.cn}
\author{Peng Tao}
\author{Yong Wang}
\author{Zhidong Zhang}
\affiliation{Shenyang National Laboratory for Materials Science,
Institute of Metal Research and International Centre for Materials Physics,
Chinese Academy of Sciences, 72 Wenhua Road, Shenyang 110016, PRC}

\date{\today}

%---------------------------------------------------------------------
\begin{abstract}
We systematically study the effect of high pressure on the structure, electronic structure and transport properties of 2H-MoS$_2$, based on first-principles density functional calculations and the Boltzmann transport theory. Our calculation shows a vanishing anisotropy in the rate of structural change at around 25 GPa, in agreement with the experimental data. A conversion from van der Waals(vdW) to covalent-like bonding is seen. Concurrently, a transition from semiconductor to metal occurs at 25 GPa from band structure calculation. Our transport calculations also find pressure-enhanced electrical conductivities and significant values of the thermoelectric figure of merit over a wide temperature range. Our study supplies a new route to improve the thermoelectric performance of MoS$_2$ and of other transition metal dichalcogenides by applying hydrostatic pressure.
\end{abstract}
%---------------------------------------------------------------------
\pacs{
31.15.A-,
% Ab initio calculations
81.40.Vw,
% High-pressure effects on structural properties of materials,
73.50.Lw,
% Thermoelectric effects
72.80.Ga
% Transition-metal compounds
}
\maketitle
%---------------------------------------------------------------------
\section{Introduction}
Molybdenite, MoS$_2$, a typical member of the transition-metal chalcogenide family, consists of alternating sandwiched sub-layers bonded by van der Waals (vdW) interaction. This weak inter-layer interaction, which has rendered it a good solid lubricant \cite{PhysRevB.48.10583}, proves also essential in determining the effect of the number of stacking layers on manifold novel properties, such as Raman frequency \cite{ACSNano.4.2695, PhysRevB.84.155413, RSC}, band-gap type (direct, indirect) and band-gap size \cite{PhysRevB.84.045409, PhysRevB.64.205416}.

Van der Waals interaction may enhance some intriguing properties on one hand, while suppressing electrical transport properties \cite{PPSB1953, JPCS1983, SSC1986, kappa2} on the other. MoS$_2$ and other transition metal dichalcogenide materials were reported to have very pronounced values of thermopower, but poor electrical conductivity \cite{PPSB1953, JPCS1983, SSC1986, kappa2}, leading to a negligibly small value of the dimensionless thermoelectric figure of merit (ZT $\sim$ 0.006 at the optimal doping and working temperature) \cite{hhguo}. To improve upon its electrical conductivity for a better value of ZT, pressure/stress may be employed to tune the inter-layer interaction and expected to suppress vdW bonding or to convert it to covalent or metallic type. Encouragingly, similar studies have already been performed on the single layer. Stress-induced enhancement of electron mobility in MoS$_2$ single layer has been demonstrated in experiment \cite{nnature.6.147} for a promising nano-electronic device. First-principles calculations have shown that both compressive \cite{MoS2-presure-theo} and tensile strain \cite{MoS2-presure-theo2} in a single layer give rise to a reduced band gap. A semiconductor-metal transition was obtained for a compressive strain of about 15$\%$ for single-layer MoS$_2$ \cite{MoS2-presure-theo2}. For bulk transition-metal chalcogenide, except for some experimental studies on pressure-induced structural change\cite{MoS2-presure-exp,MoS2-presure-exp2,MoS2-presure-exp3,MoS2-presure-exp4}, there have been so far no theoretical reports on the pressure effect. In particular, how pressure affects the electronic structure and if it can ultimately improve its thermoelectric properties have never been investigated.

In this work, combining first-principles density functional calculations with semiclassical Boltzmann transport theory, we study the change of unit cell allowed by the crystal structure, electronic structure and thermoelectric transport properties of MoS$_2$ under hydrostatic pressure. Our calculation shows a vanishing anisotropy in the rate of structural change at around 25 GPa, in agreement with the experimental data \cite{MoS2-presure-exp}. We also find a concurrent conversion from vdW interaction to covalent-like type and a semiconductor-metal transition at about 25 GPa. This transition improves greatly the inter-layer electrical conductivity but still keeps a large value of thermopower ($\sim 200 \mu V/K$), therefore giving rise to appreciably enhanced thermoelectric transport properties. Values of figure of merit (ZT) as high as 0.65 along cross-plane direction are obtainable in a wide range of temperature (from 200 K to 700 K). Our study demonstrates that applying high hydrostatic pressure can be an effective way to improve the thermoelectric performance of MoS$_2$ and of other transition-metal dichalcogenides.

\section{Methodology}
\begin{figure}[t]
\includegraphics[width=1.0\columnwidth]{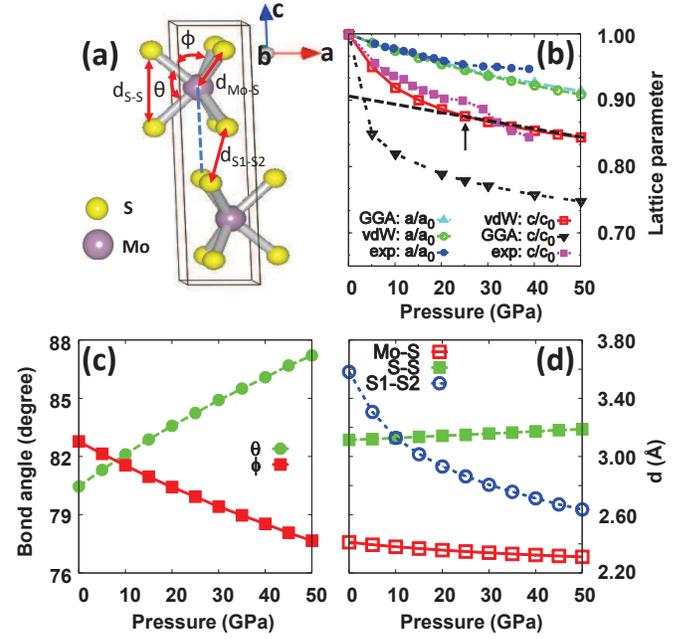}
\caption{(Color online) (a) Illustration of unit cell of MoS$_2$ and definition of the bond length d$_{Mo-S}$, d$_{S-S}$, d$_{S1-S2}$ and angles $\theta$ and $\phi$. The dependence on hydrostatic pressure of (b) lattice parameters a/a$_0$ and c/c$_0$, (c) bond angles, (d) bond lengths. Lattice parameters a/a$_0$ and c/c$_0$ in (b) are compared between GGA, GGA-vdW calculations and experiment \cite{MoS2-presure-exp}, showing GGA-vdW results agree with experiment much better than GGA. Inter-layer bonding character is likely to change around 25 GPa, as indicated by a black dashed line and arrow in (b).
\label{Fig-struc} }
\end{figure}

MoS$_2$ has P$_{63}$/mmc space group symmetry and consists of a hexagonal plane of Mo atoms sandwiched by two hexagonal planes of S atoms. The unit cell contains two alternating layers with an AB stacking along the c axis, as illustrated in Fig.~\ref{Fig-struc}(a).

We carried out first-principles calculations based on density functional theory as implemented in the VASP (Vienna ab-initio simulation package)\cite{VASP1, VASP2} within the framework of the PAW (projector augmented wave) method\cite{PAW}. The electronic exchange-correlation is described within the generalized gradient approximation (GGA) of Perdew-Burke-Ernzerhof (PBE) flavor\cite{PhysRevLett.77.3865}. The vdW corrections \cite{vdw1,vdw2} are included to determine the inter-layer spacing in 2H-MoS$_2$ and compared to the calculations without vdW corrections. A plane-wave basis set with kinetic energy cutoff of 450 eV is used. For Brillouin-zone(BZ) integrations, We use 42$\times$42$\times$10 k mesh in the full BZ. The unit cell volume, shape and internal atomic positions are optimized under each hydrostatic pressure by using the conjugate gradient method, till the internal atomic force is less than 10$^{-2}$ eV/$\AA$ and three diagonal terms in the stress matrix are equal. The convergence for energy is chosen as 10$^{-7}$ eV between two consecutive steps. We consider the external hydrostatic pressure no larger than 50 GPa, where no occurrence of structural phase transition has been observed experimentally \cite{MoS2-presure-exp} and the structure remains to have P$_{63}$/mmc space group symmetry.

Transport properties were calculated based on the Boltzmann transport theory applied to the band structure. The integration is done within the BOLTZTRAP transport code \cite{PhysRevB.68.125212}. A very dense mesh with up to 18000 k points in the BZ is used. The electron scattering time $\tau$ is assumed to be independent of energy, an approximation proved to give a good description of thermopower in a number of thermoelectric materials \cite{PhysRevB.83.115110, PhysRevB.83.245111, PhysRevB.80.075117}.

%---------------------------------------------------------------------
%---------------------------------------------------------------------

\section{Results and discussion}

%---------------------------------------------------------------------
\begin{figure}[t]
\includegraphics[width=1.0\columnwidth]{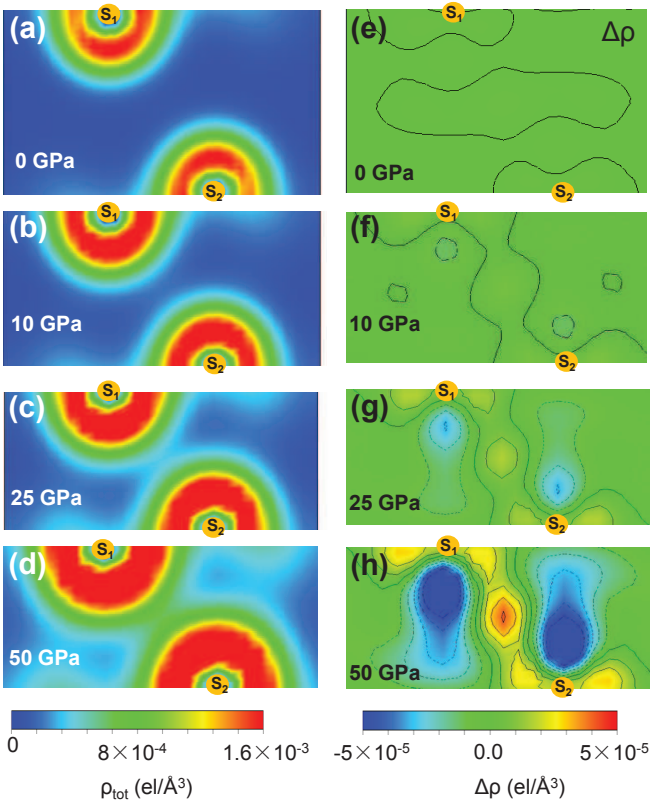}
\caption{(Color online) Total charge density $\rho$ and density of charge transfer $\Delta \rho$ of MoS$_2$ in the region between S$_1$ and S$_2$ atoms (a,e) at 0 GPa, (b,f) at 10 GPa, (c,g) at 25 GPa and (d,h) at 50 GPa, seen from (110) cutting plane. A gradual change of bonding from (a,e) van-der-Waals type to (d,h) covalent-like type is seen. $\Delta$$\rho$ = $\rho$(AB) - $\rho$(A) - $\rho$(B), with $\rho$(AB), $\rho$(A) and $\rho$(B) as the charge density of AB-stacking, of only A and of only B sub-layer, respectively. Values of charge density shown in the color bars are taken from values at 50 GPa.
\label{Fig-rho}}
\end{figure}
%---------------------------------------------------------------------
We first studied pressure effect on structure of MoS$_2$ and understand the discontinuous structural variation observed in experiment \cite{MoS2-presure-exp}. The experimental results, given in Fig.\ref{Fig-struc}(b) by filled circles and squares, show that below 25 GPa the rate of reduction of lattice parameters with pressure is different along in-plane and cross-plane directions: c/c$_0$ shrinks much faster than a/a$_0$. This anisotropy vanishes above 25 GPa. We optimized structures both with and without a consideration of the inter-layer van der Waals interaction. The calculated in-plane lattice parameters from both methods are similar and agree with experimental data, while the inter-layer parameters depend on the method. The vdW correction seems essential to account for the experiment. With vdW correction, we also calculated internal structural parameters, including bond angles and bond lengths. As shown in Fig.~\ref{Fig-struc}(c,d), the intra-layer parameters (the S-Mo-S bond angles $\theta$, $\phi$ and the bond lengths d$_{Mo-S}$, d$_{S-S}$) change very little compared to the inter-layer d$_{S1-S2}$(the distance between sulfur ions in adjacent layers). d$_{S1-S2}$, presented in Fig.~\ref{Fig-struc}(d), shrinks abruptly with pressure especially below 25 GPa. This indicates that a substantial pressure induces a change of inter-layer interaction.

%---------------------------------------------------------------------
\begin{figure}[t]
\includegraphics[width=1.0\columnwidth]{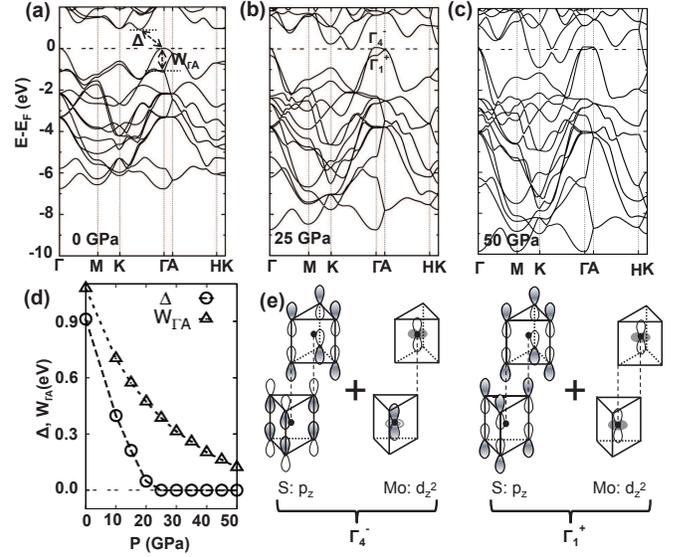}
\caption{(a-e) GGA-vdW electronic band structure of MoS$_2$ under different hydrostatic pressures. Band gaps $\Delta$ and band width W$_{\Gamma A}$ are defined in (a) and plotted in (d) as a function of pressure. The electronic states close to E$_F$ are mainly from Mo-d and S-p orbitals \cite{PhysRevB.8.3719}. Possible combinations of S-p$_z$ and Mo-d$_{z^2}$ orbitals in MoS$_2$ unit cell is illustrated in (e) with symmetry labels ($\Gamma_1^+$ and $\Gamma_4^-$ also shown in (b)) used for the irreducible representations of the P6$_3$/mmc space group (Ref. 31).
\label{Fig-band} }
\end{figure}
%---------------------------------------------------------------------

To see what happens to the inter-layer interaction upon pressure, we studied total charge densities in the region between S$_1$ and S$_2$ atom under hydrostatic pressure ranging from 0 to 50 GPa (Fig.~\ref{Fig-rho}). Compared to no charge distribution between two neighboring layers at 0 and 10 GPa in Fig.~\ref{Fig-rho}(a, b), electron charge becomes more and more converged as illustrated from Fig.~\ref{Fig-rho}(c) to (d). As d$_{S_1-S_2}$ is decreasing with pressure, the vdW interaction seems firstly to increase, due to both nonspherical-charge-distribution-induced electric dipole on S$_1$ and S$_2$ atom (Fig.2(a) and (b)) and reduced atomic distance by pressure. This trend is consistent with that in ultra-thin MoS$_2$ (few-layers)\cite{RSC}. Then a weak covalent-like bond starts to form at 25 GPa and grows stronger at 50 GPa. To support this view, we also studied charge transfer between AB-stacking charge density $\rho$(AB) and the sum of the two independent densities ($\rho$(A) + $\rho$(B)) of A and B sub-layers. There is charge transfer induced by pressure and charge accumulates mostly between S$_1$ and S$_2$ atoms.

The change of inter-layer interaction will naturally lead to dependence of band structure on pressure. Fig.~\ref{Fig-band} gives the calculated band structure evolving with pressure from 0 to 50 GPa. Fermi energies of semiconductors are set at valence band maximum (VBM). The behavior of the band gap is shown from (a) to (c) and summarized in (d). At zero pressure the indirect band gap $\Delta$ is between VBM at the $\Gamma$ point and conduction band minimum (CBM) between $\Gamma$ and K points. Applied pressure causes the CBM to drop, $\Delta$ decreases monotonically with pressure till a semiconductor-metal transition occurs at 25 GPa. This only takes an in-plane compressive strain of 5.5\%, in contrast to 15\% in single layer \cite{MoS2-presure-theo2}. It seems much easier for the bulk material than the single layer to close the band gap $\Delta$ by applying pressure, presumably due to the existing inter-layer interaction.

It is worth noting that an increase of band dispersion with pressure is observed, except for two bands near the Fermi level, between the $\Gamma$ and A points. These two bands are actually zone-folded from one band as the unit cell has one sub-layer. For convenience, both bands are treated as one with band width W$_{\Gamma A}$, as defined in Fig.~\ref{Fig-band}(a). From Fig.~\ref{Fig-band}(d), W$_{\Gamma A}$ decreases from 1.1 eV (at 0 GPa) via 0.39 eV (at 25 GPa) to 0.12 eV (at 50 GPa). To understand the counterintuitive behavior of the band dispersion decreasing with pressure, we look into the Mo-d and S-p orbitals close to the Fermi level, since they play predominate roles in forming conduction channels for both in-plane and cross-plane directions\cite{PhysRevB.8.3719}. The two cross-plane bands with the anomalous width W$_{\Gamma A}$ are due to hybridization of Mo-d$_{3z^2-r^2}$ (or d$_{z^2}$ for simplicity) and S-p$_z$ orbitals. The symmetries of possible combinations of Mo-d$_{z^2}$ and S-p$_z$ orbitals are illustrated in Fig.~\ref{Fig-band}(e). The VBM at $\Gamma$ point has a $\Gamma^{-}_{4}$ symmetry \cite{PhysRevB.64.205416, PhysRevB.8.3719, gamma4}, and the band below has a dominating contribution of S-p$_z$ orbitals with $\Gamma^{+}_{1}$ symmetry \cite{PhysRevB.8.3719, gamma1}. We found the valence band have S-p$_z$ anti-bonding characteristic. In this sense, the shrinkage of inter-layer distance d$_{S1-S2}$ with pressure suggests a stronger anti-bonding state and therefore a narrower band dispersion near E$_F$ from $\Gamma$ to A.

From the band structures shown above, we can extract some useful information on the electrical conductivity of the metallic states of MoS$_2$. In Fig.~\ref{Fig-band}(b,c), we found the Fermi velocity (v$_F$) and effective mass (m$^*$) respectively increasing and decreasing with pressure. Calculated electronic densities of states (DOSs) (not shown here) of the Fermi level at 25 and 50 GPa are 0.19 and 0.53 electrons/eV/u.c., respectively. Under thermal excitation within a range of interest (e.g., 100-700 K, equivalent to 0.009-0.060 eV of energy), DOSs translate to carrier concentrations of approximately 1.5$\times$10$^{20}$ and 5$\times$10$^{20}$ cm$^{-3}$ for 25 and 50 GPa, respectively. The boost of carrier concentration by pressure, together with increased v$_F$ and reduced m$^*$, suggest pressure-enhanced electrical conductivity. For semiconducting states below 25 GPa, pressure reduces the band gap size, which helps improve the electrical conductivity but very little. Electron/hole doping \cite{PPSB1953} was used but it turns out to be quite challenging to dope in bulk MoS$_2$. So far the highest experimental doping level on MoS$_2$ has reached only a few 10$^{16}$ cm$^{-3}$, far below 10$^{19}$ cm$^{-3}$ usually for optimizing the thermoelectric performance of semiconducting materials \cite{PhysRevB.83.115110, hhguo}.

Information on thermopower can also be extracted from the band structures. According to the Mott relation \cite{PhysRev.181.1336}, thermopower depends crucially on the derivative of the logarithmic electrical conductivity at the Fermi level \cite{jziman,PhysRevLett.104.176601}, i.e., S $\sim$$\frac{\partial \ln \sigma}{\partial E}$$|_{E=E_F}$. Based upon this formula, a strong thermopower is anticipated for both in-plane and cross-plane directions at 25 GPa. Moreover, the anomalous $\Gamma$-A bands upon pressure get more flat due to inter-layer anti-bonding states existing and its band edges locate near the Fermi level. It is therefore more advantageous for having high values of thermopower along cross-plane direction as pressure goes beyond 25 GPa.

We have discussed qualitatively how the effect of pressure on thermopower and electrical conductivity can be estimated from some aspects of the band structure. Now we turn to semi-classical Boltzmann transport theory for quantitative prediction on the thermoelectric parameters of MoS$_2$ under pressure.

\begin{figure}[t]
\includegraphics[width=1.0\columnwidth]{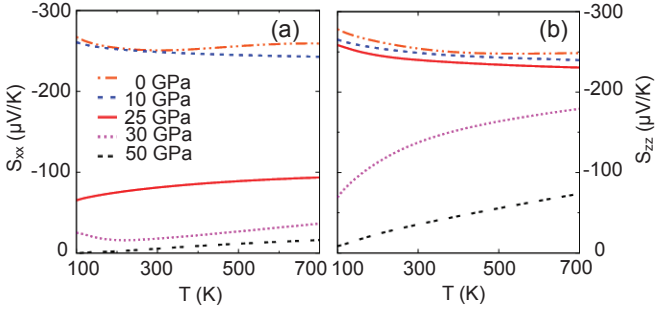}
\caption{(Color online) Dependence of thermopower on the temperature under different hydrostatic pressures along (a) in-plane direction and (b) cross-plane direction.
\label{Fig-ST} }
\end{figure}

Firstly, we calculated thermopower as function of temperature at 0, 10, 25, 30 and 50 GPa. From Fig.~\ref{Fig-ST}, a strong anisotropy of the thermopower between two perpendicular directions appears when pressure increases from 0 to 25 GPa. The in-plane thermopower S$_{xx}$ drops substantially while the cross-plane thermopower S$_{zz}$ goes down fairly modestly. The S$_{xx}$ at pressure above 25 GPa is as low as in typical metals\cite{Blattbook}. In the contrary, high values of the Seebeck coefficient S$_{zz}$ were indeed obtained along cross plane direction. Especially, over 100 $\mu$V/K along c-axis are achievable over a wide temperature range above 25 GPa. Least important to mention is that a big value of thermopower is also found for the semiconducting MoS$_2$ at 0 and 10 GPa (with experimental doping $\sim$ 10$^{16}$ cm$^{-3}$), which is however not very unusual from theoretical perspective \cite{mahan1998}.

%---------------------------------------------------------------------
%---------------------------------------------------------------------

\begin{figure}[t]
\includegraphics[width=1.0\columnwidth]{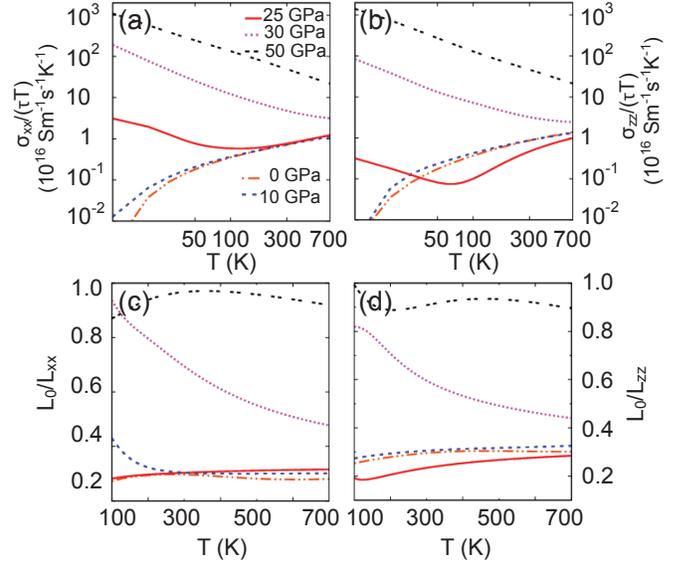}
\caption{(Color online) $\frac{\sigma}{\tau T}$ as function of pressure and temperature for (a) in-plane and (b) cross-plane direction. The ratio of Lorenz number L$_0$ over $\frac{\kappa_e}{\sigma T}$ along (c) in-plane and (d) cross-plane direction is given to show how conductive it is under various pressures from 0 GPa to 50 GPa in the background of the Wiedemann-Franz law.
\label{Fig-lorenz} }
\end{figure}

Similarly, we then calculated $\sigma$/$\tau$ as a function of temperature. Our results show that $\sigma$/$\tau$ increases with external pressure. So does the electrical conductivity $\sigma$, if $\tau$ is approximately pressure independent. To find a realistic temperature dependence of $\sigma$ without available experimental $\tau$, we assume that at high temperature the conduction electrons are mainly scattered by phonons and the electronic relaxation time $\tau$ $\sim$ T$^{-1}$ is used \cite{jziman, jziman2}. Hence $\sigma$/($\tau$T) as function of temperature can give information on temperature dependence of $\sigma$. We show in-plane and cross-plane $\sigma$/($\tau$T) versus temperature in Fig.\ref{Fig-lorenz}(a,b). For both directions, typical metallic behavior is seen by an inverse power law relation between $\sigma$/($\tau$T) and temperature at pressure higher than 25 GPa, while below 25 GPa it is semiconducting. At 25 GPa, a deviation from the inverse power law is observed and suggests its conductive properties different from typical metals and semiconductors.

%---------------------------------------------------------------------
%---------------------------------------------------------------------

%---------------------------------------------------------------------
\begin{figure}[b] \includegraphics[width=1.0\columnwidth]{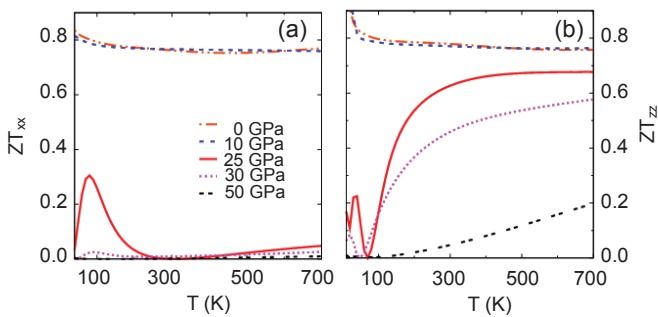}
\caption{(Color online) Temperature and pressure dependence of figure of merit ZT of MoS$_2$ along (a) in-plane and (b) cross-plane directions. The cross-plane is more preferable for practical thermoelectric application.
\label{Fig-ZT-T} }
\end{figure}
%---------------------------------------------------------------------

As is known, the dimensionless thermoelectric figure of merit ZT is defined as $\frac{S^2 \sigma T}{\kappa}$ (or $\frac{S^2}{L + \eta}$), with $\kappa$ (=$\kappa_e$ + $\kappa_l$), $\kappa_e$ and $\kappa_l$ as total, electronic and lattice thermal conductivity, respectively, L = $\frac{\kappa_e}{\sigma T}$ and $\eta$ = $\frac{\kappa_l}{\sigma T}$. To evaluate a possible enhancement of ZT value by external pressure, we need to calculate the pressure dependence of thermal conductivity $\kappa$. Unfortunately, calculating lattice thermal conductivity $\kappa_{l}$ is far more complicated than calculating $\kappa_e$ and it is beyond band structure calculations. Here we argue that $\kappa_e$ can substitute for $\kappa$ under various pressures if MoS$_2$ becomes metallic, since $\kappa_l$ is insignificant compared to $\kappa_e$ in most metals\cite{kappalatt}. For a typical metal, the Wiedemann-Franz law works and L is a constant called Lorenz number L$_0$(= 2.8 $\times$ 10$^{-8}$ W$\Omega$/K$^2$). In Fig.~\ref{Fig-lorenz}(c,d), we draw a ratio of L$_0$ over L as function of temperature at various pressures. The ratio L$_0$/L approaches unity as pressure increases from 25 to 50 GPa, indicating an evaluation of ZT value based on $\kappa_e$ is legitimate. However, a strong deviation from the Wiedemann-Franz law occurs to the semiconducting states below and at 25 GPa, signifying $\kappa_l$ may be as important as or even more significant than $\kappa_e$.

Assuming ZT is approximately $\frac{S^2 \sigma T}{\kappa_e}$ and bearing in mind that it may not work for semiconducting states even with heavy doping, we show the calculated ZT value in Fig.~\ref{Fig-ZT-T}. Though ZT values for semiconducting MoS$_2$ at 0 and 10 GPa seem practically high, it is basically unreliable since the $\kappa_l$ can be orders of magnitude higher than $\kappa_e$ and brings down ZT by similar orders of magnitude\cite{hhguo}. At transitional 25 GPa, ZT can reach 0.05 in plane and 0.65 along the c axis over a wide range of temperature. But due to the uncertainty between $\kappa_l$ and $\kappa_e$ this may need experimental confirm. What's promising here is that the values of cross-plane ZT decrease modestly as pressure goes beyond 25 GPa. Still significant ZT value higher than 0.10 is obtained over a wide range of pressure, suggesting that a proper range of pressure is essential to optimize the thermoelectric properties of MoS$_2$.

\section{Conclusion}
In summary, we systematically studied the structure, electronic structure and transport properties of 2H-MoS$_2$ under hydrostatic pressure, based on the Boltzmann transport theory and first-principles density functional calculations. VdW corrections to GGA calculations are essential to deal with such an anisotropic system under pressure. Our calculation shows a vanishing anisotropy in the rate of structural change at around 25 GPa, in agreement with the experimental data. A conversion from vdW to covalent-like bonding is seen. Concurrently, a transition from semiconductor to metal occurs at 25 GPa. A much smaller compressive strain in bulk material than in single layer is needed to close the band gap. We obtained in the metallic state very large values of the thermopower and ascribed it to some anomalous band features. Above all, good values of figure of merit are found along the cross-plane direction. The ZT value can be obtained as big as 0.10 (even up to 0.65) over a wide pressure and temperature range. Our study supplies a new route to improve the thermoelectric performance of MoS$_2$ and of other transition metal dichalcogenides by applying hydrostatic pressure.

\begin{acknowledgments}
The authors thank Dr. Peter F. de Ch$\hat{a}$tel for useful discussions. This work was supported by the NSFC under Grant No. 11004201, 50831006 and the National Basic Research Program (No. 2012CB933103). T.Y. acknowledges IMR SYNL Young Merit Scholars and T.S. K$\hat e$ Research Grant for support.
\end{acknowledgments}

%\bibliographystyle{apsrev}% your bst file here
%\bibliography{TE-MoS2} %your bib file here

\begin{thebibliography}{38}
\bibitem[{kap()}]{kappalatt}
\bibinfo{note}{The thermal conductivity $\kappa_l$ is insignificant compared to
  $\kappa_e$ in metallic states, due to the following reasons:(1) At low T,
  $\kappa_e$ dominates in the metals. (2) At high T, $\kappa_e$ is
  temperature-insensitive under circumstance of $\tau$ $\sim$ T$^{-1}$, while
  $\kappa_l$ decreases with temperature, as experiments suggested
  \cite{kappa2}. It makes $\kappa_l$ even more insignificant than $\kappa_e$.
  (3) It would reduce our estimation of ZT by only one half if $\kappa_l$ were
  comparable to $\kappa_e$.}
\end{thebibliography}

\end{document}